\documentclass{ieeeaccess}
\usepackage{cite}
\usepackage{amsmath,amssymb,amsfonts}
\usepackage{algorithmic}
\usepackage{graphicx}
\usepackage{textcomp}

\usepackage{multirow}

\usepackage{caption,setspace}
\def\BibTeX{{\rm B\kern-.05em{\sc i\kern-.025em b}\kern-.08em
    T\kern-.1667em\lower.7ex\hbox{E}\kern-.125emX}}
\begin{document}
\history{Date of publication xxxx 00, 0000, date of current version xxxx 00, 0000.}
\doi{10.1109/ACCESS.2017.DOI}

% \title{Advancing Cognitive Training with VR: A Framework for Prospective Memory Enhancement}
\title{Preliminary Study on Virtual Reality Framework for Effective Prospective Memory Training: Integration of Visual Imagery and Daily-life Simulations}

\author{
% \uppercase{Satoshi Fukumori}\authorrefmark{1}, \IEEEmembership{Member, IEEE},
\uppercase{Satoshi Fukumori}\authorrefmark{1},
\uppercase{Kayoko Miura\authorrefmark{2}, Ayako Takamori\authorrefmark{3} and Sadao Otsuka}.\authorrefmark{4}}

\address[1]{Department of Design and Engineering, Kagawa University, Saiwai-cho 1-1, Takamatsu City, Kagawa prefecture Japan}
\address[2]{Nagasaki Junshin Catholic University, Mitsuyama-cho 235, Nagasaki city, Nagasaki prefecture Japan}
\address[3]{Clinical Research Center, Saga University Hospital, Nabeshima 5-1-1, Saga city, Sagaprefecture, Japan}
\address[4]{Department of Clinical Psychology, Graduate School of Education, Hyogo University of Teacher Education, 1-5-7, Higashikawasaki-cho, Chuo-ku, Kobe, Hyogo, Japan and Department of Psychiatry, Graduate School of Medicine, Kyoto University, 54 Shogoin-kawahara-cho, Sakyo-ku, Kyoto, Japan}
\tfootnote{This work was supported in part by JSPS KAKENHI Grant Number 20H01776.}

\markboth
{Author \headeretal: Preparation of Papers for IEEE TRANSACTIONS and JOURNALS}
{Author \headeretal: Preparation of Papers for IEEE TRANSACTIONS and JOURNALS}

\corresp{Corresponding author: First A. Author (e-mail: fukumori.satoshi@kagawa-u.ac.jp).}

\begin{abstract}
Prospective memory (PM), defining the currently conceived intention of a future action, is crucial for daily functioning, particularly in aging populations. 
This study develops and validates a virtual reality prospective memory training (VR--PMT) system that integrates visual imagery training (VIT) and virtual reality training (VRT) to enhance the PM abilities of users. 
The framework is designed to progressively challenge users by simulating real-life PM tasks in a controlled VR environment. 
The VIT component is designed to improve the generation and utilization of visual imagery by users, while the VRT component provides PM tasks based on time and event cues within a virtual environment.
The framework was evaluated on ten healthy adults (university students and elderly participants) over nine weeks. 
During the initial session, the baseline PM abilities of the participants were assessed using the memory for intentions screening test (MIST). 
The subsequent sessions alternated between VIT and VRT with increasing task complexity. 
The MIST scores were significantly positively correlated with task achievement, confirming the efficacy of the system. Imagery abilities were also strongly correlated with task performance, underscoring the importance of visual imagery in PM training.
Usability and user experiences, evaluated on the Jikaku-sho Shirabe questionnaire and the user experience questionnaire, indicated an overall positive user experience but higher fatigue levels in elderly participants. 
This study demonstrates that the VR--PMT system effectively trains and assesses PM abilities by integrating VIT and VRT, supporting its potential for broader applications in clinical settings. 
\end{abstract}

\begin{keywords}
Prospective memory, Virtual reality, framework
\end{keywords}

\titlepgskip=-15pt

\maketitle

\section{Introduction}
\label{sec:introduction}

\PARstart{P}{rospective} memory (PM), defining the ability to remember and perform an intended action at a specific future time or in a particular future context\cite{PM_Martin_2003},
plays a crucial role in the daily lives of humans. For example, PM initiates actions such as turning off the stove to prevent a possible fire or taking medication at a specified time to avoid potential health consequences. 
PM is important for maintaining the quality of life\cite{Zogg_2012, Woods_2012, Woods_2014, Woods_2015, Sheppard_2020} and for independently performing daily activities\cite{Supporting_PM_Zollig_2007}. 
PM failure is a serious problem for family members and caregivers, who are required to remember the schedule of the affected person. 
Previous studies have indicated that PM ability declines with age. In particular, older adults perform less well on PM tasks than young adults\cite{Henry_2004, Kvavilashvili_2009, Koo_2021}. 
PM ability is also altered by mental disorders such as brain injuries, schizophrenia, and Alzheimer's disease \cite{Wang_2009, Berg_2012, Raskin_2020}.

Psychological interventions targeting PM impairment, such as PM training, have been extensively studied. 
Such interventions have proven their effectiveness in various populations, including older adults and individuals affected by various diseases \cite{Jones_2021, Hering_2014}. As highlighted by
Raskin\cite{Visual_imagery_Raskin_2019}, visual imagery training, a technique for visualization from a given word, is an important self-supporting strategy that can potentially improve PM deficits.
Umeda\cite{umeda_2014_mini_day_task} attempted to improve PM impairment in brain injury patients through training on Mini-day tasks. 
They instructed participants to visualize themselves performing actions in real-life scenarios while remembering the tasks, thereby connecting the intervention to everyday situations. 
Mathews\cite{Mathews_2016} reported that PM impairment in brain injury patients was improved after cognitive training incorporating virtual reality (VR) technology that combines strategy acquisition with practical application. 
Training based on visual imagery strategies (the internal strategy approach) can be enhanced through VR technology. 
Moreover, head-mounted displays can resolve a major challenge in conventional PM studies, namely, minimizing the environmental differences between laboratory and real-life settings. 
% However, Mathews' study was performed on a small sample size of 15 participants and did not include a control group. 
% Many previous studies on the effectiveness of PM training have utilized similarly small sample sizes without control groups. Such studies rely heavily on case reports or pre- and post-intervention comparisons. 
% A VR training system with a more systematic training program would allow a more comprehensive investigation.

Many training methods that aim to improve cognitive functions are limited by the differences between real-world environments and laboratory settings. 
Several studies\cite{Valmaggia_2016, Calafell_2018} have indicated that VR methodologies can recreate social events within a controlled laboratory environment.
As traditional training methods related to PM cannot fully replicate the PM scenarios encountered in daily life, their training effectiveness is limited. 
In contrast, VR technology can recreate realistic environments, allows users to immerse themselves in the training process. 
VR can also simulate daily life scenarios within a laboratory setting, improving the effectiveness of PM training. 
Huang\cite{Huang_2022} demonstrated the usefulness of VR in PM training systems for patients with depressive disorders. 
Hogan\cite{Hogan_2023} reported that VR also assists PM in stroke patients. 
However, a systematic and effective framework for constructing VR training systems is currently lacking.

% In the present study, we propose and implement a framework that creates a VR training system for PM. 
% % In the present study, we propose and implement a framework to develop a VR training system for enhancing PM skills. This framework includes , task types and their selection, difficulty levels, and training environments. 
% Although intended for individuals with PM problems, the system is preliminarily validated on healthy individuals to ensure that it accurately reflects PM capabilities. 
% If the system is properly designed, the task achievement rates should reflect the PM abilities of the participants, confirming that the system effectively measures and trains PM.
In this study, we propose and implement a framework to develop a VR-based training system for enhancing PM skills. 
This framework comprises task types, difficulty levels, and realistic training environments to simulate daily PM challenges. 
The effectiveness of the proposed framework is evaluated by implementing the training system and conducting a preliminary validation with healthy  adults who do not have clinical dementia to analyze the system’s capacity to accurately reflect PM abilities. 
If properly designed, the system should indicate task achievement rates corresponding to the PM capabilities of participants.
\section{Related work}
\label{sec:Related_work}

Training and evaluating PM abilities is an inherently challenging task. 
Several tests\cite{RPA_ProMem, CAMPROMPT, MIST} have been established for assessing PM capabilities, but the evaluated tasks are limited to simple and abstract activities. 
For instance, the Cambridge Prospective Memory Test (CAMPROMPT) provides six pen-and-paper tasks for PM assessment. 
As PM is intricately linked to real-life environments and specific times of the day, abstract tasks are insufficient for a comprehensive evaluation. Recall can be aided by the passage of time (e.g., by the approach of evening, reminding the individual that shopping time is 5 PM) and environmental cues (e.g., a telephone in view that reminds the individual of their intention to make a phone call). 
By simulating real-world environments and times, VR can create realistic settings, schedules, and tasks. 
Such VR-created environments offer highly interactive modalities and can achieve high ecological validity\cite{Kourtesis_2020_VR_EAL,Kourtesis_2021_VR_EAL}. 
Various VR-based evaluations of PM in patients with brain injuries have demonstrated the potential for VR in creating realistic and effective assessment environments\cite{Brooks_2004, Knight_2009}.

PM requires both remembering and recall abilities. Few specialist training methods for PM can simulate real-life situations, although 
several traditional training methods can create these conditions. 
In Virtual Week\cite{Virtual_week_2009, Mioni_2015}, a board game-style training, the
participants move around the board using dice, select plausible daily activities, and must remember and complete PM tasks. 
Umeda's Mini-day task\cite{umeda_2014_mini_day_task} uses cards with schedule-related contents and times (e.g., "10:00; buy a ticket at Hakata Station"). 
The trainees visualize and memorize the schedules written on the cards and then execute the schedules on an accelerated timeline. 
When applied to training methods such as Virtual Week and Mini-day tasks, VR can simulate complex and dynamic environments, providing participants with more realistic situations. 
Yip\cite{Yip_2013} developed a training program using desktop-based VR. This system provides participants with realistic situations, but the tasks are limited to shopping. 
Mathews\cite{Mathews_2016} developed a system with more complex and realistic tasks. 
This system adopts and extends Virtual Week, along with Mini-day and Mathews' system. The two latter methods provide no practice for imagery and do not account for differences among individuals' abilities to control their visual imagery; instead, imagery generation is left to the individuals' discretion\cite{Miura_2010}. 
As visual imagery and associations are challenging to some individuals, visual imagery strategies can be effectively harnessed only when tailored to individual needs. 
Therefore, the present study incorporates training into visual imagery strategies.
VR is advantageous for visual imagery because it can display a wide variety of images. 
Therefore, VR is ideally suited for incorporating an imagery strategy into a training system. 
% Accordingly, this study adopts imagery capability as a support of PM and recall.% Mathhews' system, which applies an imagery strategy to daily life activities, is divided into two phases: Imagery training and VR training in realistic situations. 
% During the imagery training phase, participants are encouraged to visualize themselves performing tasks, which has been shown to enhance PM. 
% During the second phase, the participants apply the skills acquired during the imagery training phase. 
% By incorporating both phases, Mathews' system effectively combines internal strategy training with practical application in a realistic environment.
Mathews' system\cite{Mathews_2016} integrates visual imagery strategies with VR training in realistic situations, making it particularly effective for enhancing PM. 
The system\cite{Mathews_2016} is structured into two phases. The first phase focuses on imagery training and encourages participants to visualize themselves performing tasks; this technique is known to enhance PM. 
In the second phase, participants apply these visualization skills within a realistic VR environment, bridging internal strategy training with practical application. 
This two-phase approach that combines visualization and realistic simulation adopts Mathews' integrated method for PM training.
Among many suggested subtypes of PM tasks\cite{Shum_2002, Guajardo_2000, Park_1997}, time- and event-based tasks are the most suitable for clinical tests and 
are most commonly adopted\cite{Park_1997, Kliegel_2001, Hicks_2005}. 
Time-based tasks require the individual to remember and perform an action at a specific time or after a certain period of time.
A typical time-based task is meeting somebody at a restaurant at 19:00. 
In contrast, event-based tasks is one triggered by another event, such as taking medication after meals. 
Virtual Week\cite{Rendell_2009} sets regular and irregular tasks as rule-based and non-rule-based tasks, respectively. Both types of tasks are required in VR training.
An effective training system must also provide individuals with purposeful PM tasks that they will likely encounter in daily life. 
VR-based PM training often focuses on time/event-based tasks\cite{Henry_2004, Kvavilashvili_2009, Koo_2021}. % The present paper considers PM tasks from perspectives other than time/event-based tasks, acknowledging the complexity and variety of real-life scenarios. 

The present study requires to integrate time-based/event-based and regular/irregular tasks for considering the complexity and variety of real-life scenarios.
Daily life usually includes regular schedules (e.g., taking medication after meals) and irregular schedules (e.g., going to the bank). Regular tasks establish a routine, reinforcing memory through repetition. In contrast, an irregular task, such as collecting a suit from the dry cleaners while shopping, requires flexibility and the ability to remember less frequently performed tasks. 
By including both types of tasks, the training system can better simulate the variety of prospective memory challenges encountered in daily life, thus providing a more comprehensive training experience. 
Rose\cite{Rose_2010_re_irrg} and Shelton\cite{Shelton_2016_re_irrg} involved both PM task types (regular/irregular) and PM cue types (time/event) in Virtual Week, demonstrating the importance of considering multiple dimensions of PM. 
Therefore, a VR-based training system must encompass both regular and irregular tasks, along with time-based and event-based tasks, to ensure well-rounded and effective training tasks.

\section{Design of Framework}

The VR--PMT program integrates visual imagery training (VIT) for PM task support and virtual reality training (VRT) for simulating a realistic environment. 
The VIT is designed to enhance the trainee's ability to generate and use mental images, which is an important skill in PM. 
The training is implemented by VIT followed by VRT. 

This section defines the design requirements of the VIT and VRT programs, particularly focusing on the design of the VRT program. 

\subsection{Visual imagery training}

The VIT program aims to enhance the generation and utilization of visual imagery as a PM strategy prior to the VRT program. Our program adopts the methods of Mathew\cite{Mathews_2016} and Raskin\cite{Visual_imagery_Raskin_2019}.
In VRT, a daily plan is expressed as a sentence structured with a subject (yourself), verb (action), objective, and place (e. g. "I collect my suit from the cleaners"), requiring trainees to convert words into images. 
The VIT program considers individual differences in imagery ability and emphasizes practice in imagery generation. 
The training content becomes progressively more difficult. Trainees begin with learning pairs of specific words and advance to remembering complex everyday tasks.

Next, the trainees learn visual imagery stepwise through a series of training levels in VRT. 
Structured into eight levels, the program gradually reduces support for visualization creation to increase the trainee's independence. 
In Levels 1 to 4, participants visualize pairs of nouns (such as "pumpkin and milk"). 
Levels 5 and 6 involve pairs of nouns and actions (e.g., "egg + making omelet"), and higher levels include cues and actions corresponding to time- and event-based tasks for each pair (e.g., "pass supermarket + buy strawberries").

\subsection{Virtual reality training}

% VRT also requires at least two places, i.e., a house and outside environment, to simulate a real-life scenario. 
% % VRT requires fundamental functions to create an effective virtual environment that mimics real life. % Users must remember to perform PM tasks at specific time points in relation to place or a specific event.

% The most basic function of VRT is incorporating both place and time elements. 
% The architectures and objects in the virtual environment must reflect the daily living scenarios of the users.
% In addition, distractor tasks require to simulate real life in the virtual environment. 
% To execute PM tasks effectively in the VR environment, users must know the current simulated time and track the passage of time as they would in real life. 
% In addition, the simulated time must be faster than real time to ensure efficient and manageable training.
In VRT, users must remember to perform PM tasks at specific times in relation to certain places or events. 
A foundational function of VRT is to incorporate realistic place and time elements. 
To effectively simulate real-life scenarios, VRT must have at least two settings, such as a home environment and an outdoor area, that mirror daily living conditions. 
The architecture and objects in the virtual environment should reflect familiar surroundings.
% In addition, distractor tasks should be integrated to further enhance the realism of user experience.
% Distractor tasks are alternatives of daily activities, except for PM tasks, and prevent task recalling via memory rehearsal.

For the effective execution of PM tasks, users must have access to the current simulated time and should be able to track the passage of time as they would in real life. 
Moreover, simulating an entire day in real time would impose an unrealistic cognitive load on users. Therefore, the virtual environment accelerates time to ensure efficient and manageable training. 
Previous studies (\cite{umeda_2014_mini_day_task}, \cite{Virtual_week_2000} and \cite{Mathews_2016}) have demonstrated the effectiveness of using simulated time in similar training scenarios, enabling users to experience a condensed, yet comprehensive, daily routine.
In VRT, users simulate their daily lives multiple times. 
The tasks in each simulation are designed and categorized by cue type (time-based or event-based) and by frequency (regular or irregular). 
PM tasks depend on external cues that stimulate recall. 
Event-based tasks use cues (e.g., meeting someone or having lunch) to trigger action recalls, while time-based tasks rely on specific times and the passage of time. During a time-based task, the user must check the time and act accordingly. 
Regular tasks are repeated daily, while irregular tasks are performed once only. 
Combining both types provides tasks that reflect real-life diversity. 
To help users gradually adapt to real-life situations, the task difficulty increases with each training session. 
Previous studies have shown that time-based tasks are more challenging than event-based tasks\cite{Hicks_2005}\cite{Mathews_2016}; moreover, irregular tasks are more challenging than regular ones\cite{Rose_2010_re_irrg},\cite{Shelton_2016_re_irrg}. 
Clearly, tasks combining time-based and irregular elements are the most difficult, while those combining regular and event-based elements are the easiest. 
% The initial training stages thus focus on regular and event-based tasks. More irregular and time-based tasks are introduced as the training progresses. 
% As regular tasks are required to maintain real-life scenarios, they are retained while adding new tasks that combine irregular or time-based elements.

VRT also requires distractor tasks to mimic real life in the virtual environment.
Users engage in distractor tasks when not performing scheduled PM tasks, substituting actions typically performed in daily life such as housework or work-related activities. 
Distractor tasks prevent users from rehearsing their PM tasks, promoting more natural recall of their intended actions. 
Additionally, distractor tasks should be enjoyable to engage users in the training experience. 
Each area in the training environment, such as the house or shopping area, should have at least one distractor task point. 
Including at least one distractor task point in each area of the training environment, such as the house or shopping area, ensures that the virtual environment reflects realistic daily-life conditions. 
In real life, people are often engaged in various activities (e.g., housework at home, browsing in a store) while waiting for the right time or event to execute PM tasks. 
By integrating distractor tasks into each area, the system helps simulate this natural multitasking environment, fostering better immersion and task relevance. 
Additionally, distractor tasks should be enjoyable to engage users in the training experience and prevent them from rehearsing their PM tasks, promoting more natural recall of their intended actions.

In the errorless learning method, users learn the correct responses at the beginning and repeatedly practice the correct responses, thereby minimizing their errors during the learning process\cite{errorless}. 
According to Sohlberg, errorless learning is primarily intended to reduce errors during the learning phase.
The aim of VRT is the perceived successful completion of each task. 
If users forget to execute a time-based task, the system displays a message reminding them of the task without instilling a sense of failure. The message might read "Oops, it's time for your scheduled task." If users perform an incorrect action of the intended task, they are alerted with a simple sound.

To effectively simulate real-life situations in VRT, users must familiarize themselves with the system's operations and the spatial layout of the environment, which often requires extensive pre-instruction or memorization of the map layout. 
To reduce a time for these activities, the system should provide participants can engage in tasks for VIT within the VRT environment.. 
This integration helps them familiarize themselves with the system's operations and acclimatize to the map layout. 
In addition, practicing regular and event-based PM tasks within the same environment after VIT further facilitates participants' adaptation to VRT.
This approach minimizes the cognitive demands  during VRT sessions by gradually building familiarity and reducing the effort required to remember task-related information.
\section{Implementation}

The VR--PMT system is implemented under the framework designed to integrate VIT and VRT programs. 
This section details the hardware and software setups of both programs.

\subsection{Hardware}

The VR--PMT system provides an immersive experience through a VR headset with controllers. 
For this purpose, Meta Quest 2 was chosen for its high resolution, comfortable fit, and reliable tracking system. A user receives immersive visual and auditory feedback and interact within the VR environment using hand controllers, mimicking human--environment interactions in the real world. 
The Meta Quest 2 headset is connected to an Alienware m15 R4 PC equipped with an Intel i9 processor, 32GB RAM, and an NVIDIA GTX 3080 GPU that records the behavior logs and task achievement, ensuring smooth operation and rendering of virtual environments.

\subsection{Program schedules and software setup}

Each one-hour training session was conducted once weekly. 
To ensure effective and safe training, each session was supervised by an assistant 
with multiple roles: providing support, ensuring smooth operation of the VR system, and helping the participants with any difficulties encountered during the training.
Each session of the experimental schedule focused on different aspects of VIT and VRT (see Table \ref{tab:schedule} for the contents of each session). 
% Before the first session(Sessions 0), participants completed the Memory for Intentions Screening Test (MIST) to assess their baseline PM ability. 
In Sessions 1--3, the participants engaged in the VIT program and practiced the VRT program. 
Session 4 was dedicated to a tutorial in which the assistant re-iterated the VRT program and demonstrated the regular tasks to be performed in the VRT program. 
In Sessions 5--8, the participants fully engaged in the VRT program, applying the skills learned in previous sessions.

    \begin{table}[hbtp]
    \caption{Experimental schedule}
    \centering
    % \scalebox{0.85}{
    \begin{tabular}{llll}
         \hline
        Session & \multicolumn{3}{c}{Content} \\
         & \multicolumn{1}{l}{VIT} & & \multicolumn{1}{l}{VRT} \\
         % \hline
         \cline{2-2} \cline{4-4}
         % 1 & \multicolumn{2}{l}{MIST} \\
         1 & \multicolumn{1}{l}{Lv. 1 - 3} & & \multicolumn{1}{l}{Practice} \\
         2 & \multicolumn{1}{l}{Lv. 4 - 6} & & \multicolumn{1}{l}{Practice} \\
         3 & \multicolumn{1}{l}{Lv. 7 - 8} & & \multicolumn{1}{l}{Practice} \\
         4 &                               & & \multicolumn{1}{l}{Tutorial} \\
         5 &                               & & \multicolumn{1}{l}{Lv. 1} \\
         6 &                               & & \multicolumn{1}{l}{Lv. 2} \\
         7 &                               & & \multicolumn{1}{l}{Lv. 3} \\
         8 &                               & & \multicolumn{1}{l}{Lv. 4} \\         \hline
    \end{tabular}
    \label{tab:schedule}
    \end{table}
    
The integration of VIT and VRT into the VR--PMT system ensures a seamless transition between the two phases, enhancing the overall training efficacy.
During the first session, participants received explanations about PM and visual imagery strategies and practiced operating the VR--PMT system. 
The participants first engaged in VIT and subsequently practiced operations in VRT. 
In the second session, participants continued with VIT and practiced the necessary background tasks of VRT. 
The third and fourth sessions provided the participants with further VRT practice and a tutorial on the VRT phase, respectively. 
During the tutorial, the participants listened to detailed instructions and engaged in tasks designed for the tutorial.

\subsection{VIT}

In the VIT, the participants progressed through eight levels designed to enhance their ability to generate and use mental images. 
At each level, support for creating visualizations was gradually reduced to promote independent imagery generation.
The content and expression at each level of the VIT are listed in Table \ref{tab:content_of_VIT} (where VI means visual image).
The training began with visualizing pairs of nouns and concluded with creating sentences involving actions and events.

    \begin{table}[hbtp]
    \caption{Content of VIT}
    \centering
    % \scalebox{0.85}{
    \begin{tabular}{lp{8.5mm}lp{1mm}p{8.5mm}l}
         \hline
         & \multicolumn{2}{c}{Simulation word} & & \multicolumn{2}{c}{Response word} \\
         \cline{2-3} \cline{5-6}
        Lv. & content & expression & & content & expression \\         \hline        1 & Noun & VI + word & & Noun & VI + word \\

        2 & Noun & VI + word & & Noun & VI + word \\
        3 & Noun & VI + word & & Noun & word \\
        4 & Noun & word & &  Noun & word \\
        5 & Noun & VI + word & & Action & sentence \\
        6 & Noun & word & & Action & sentence \\
        7 & Event & VI + sentence & & Action & sentence \\
        8 & Event & sentence & & Action & sentence \\
         \hline    \end{tabular}

    % }    
    \label{tab:content_of_VIT}
    \end{table}

\subsection{VRT}

\begin{figure}[tbh!]
    \centering
    \includegraphics[width=80mm]{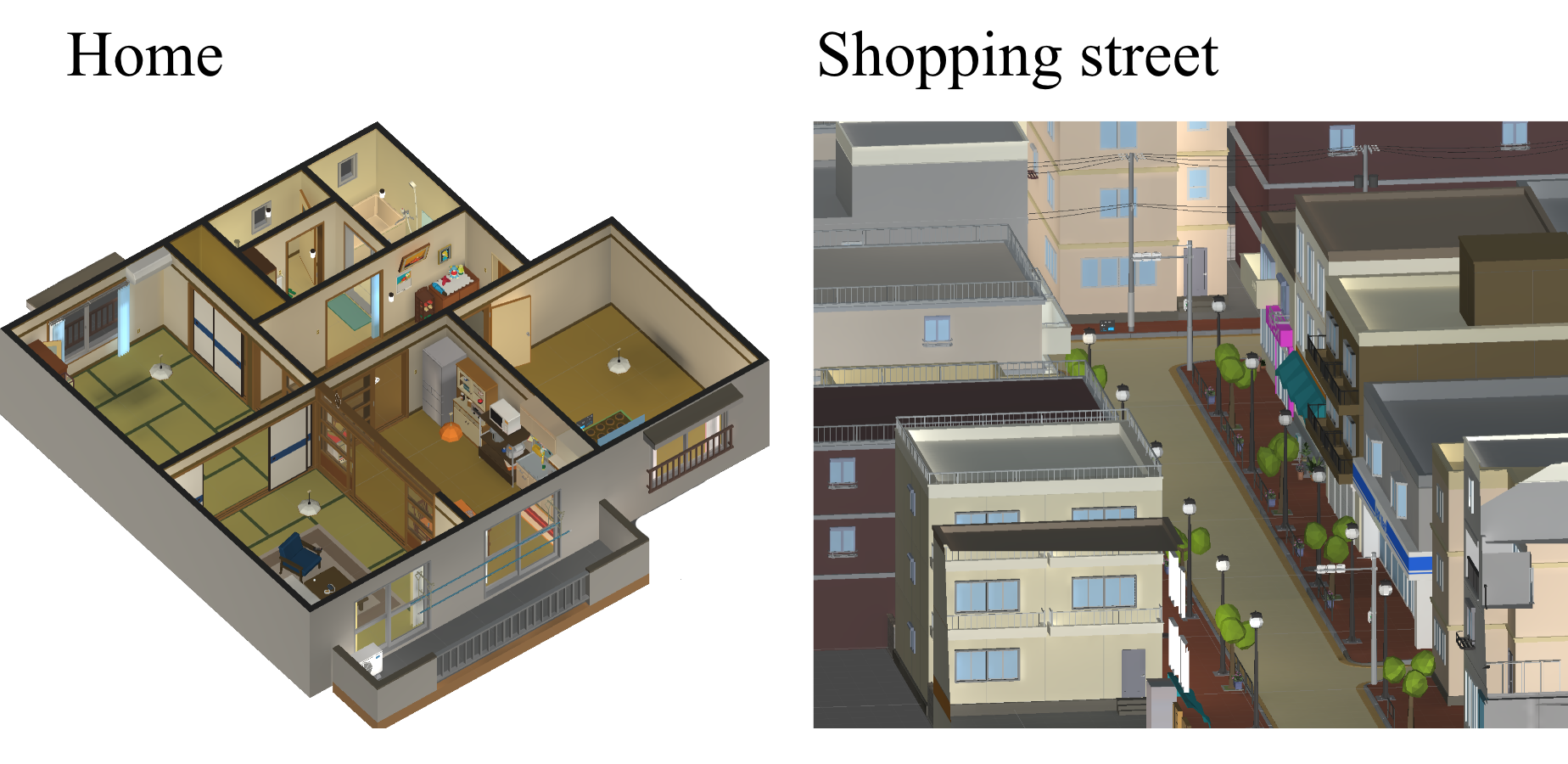}
    \caption{Layouts of the home and shopping street in the virtual environment}
    \label{fig:home_shopping_street}
\end{figure}

\begin{figure}[tbh!]
    \centering
    \includegraphics[width=80mm]{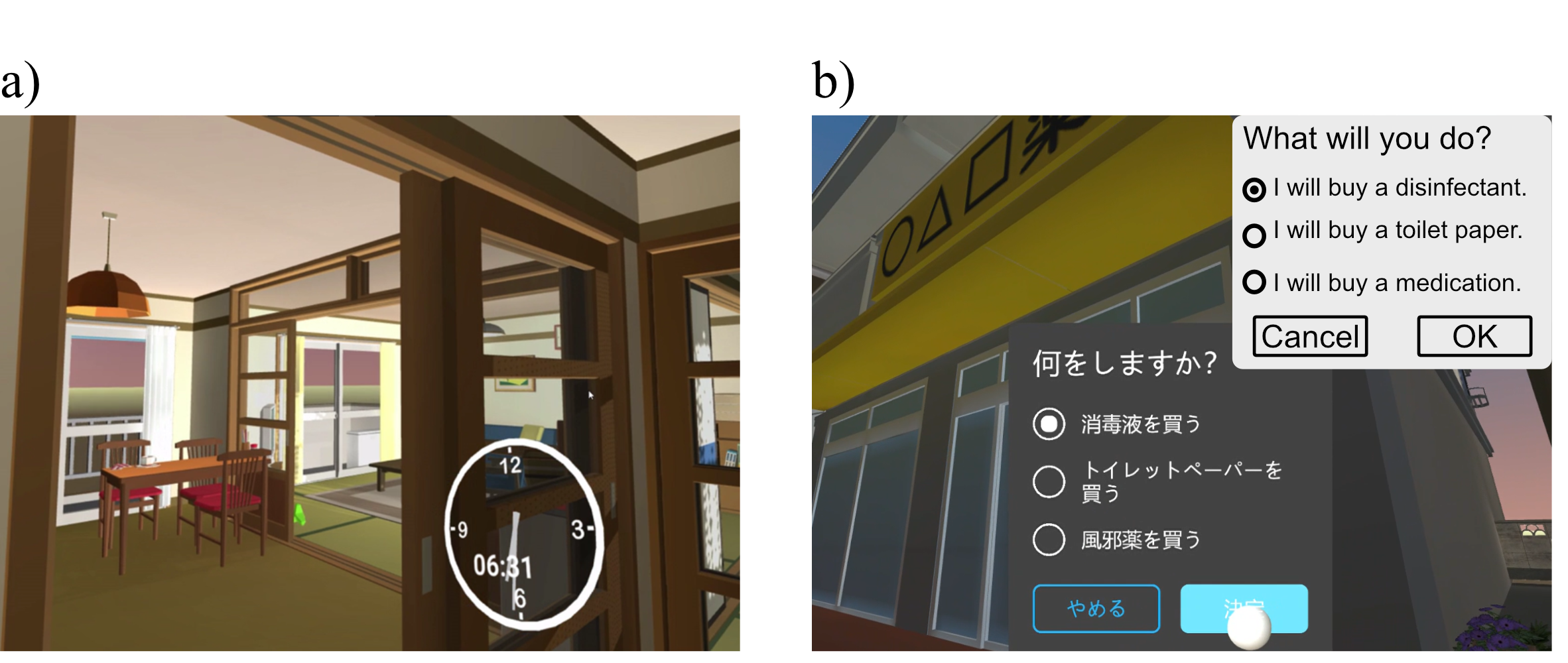}
    \caption{White clock for monitoring the time (left) and a user interface on which participants select their answer (right)}
    \label{fig:VRT_clock_task}
\end{figure}

The VR environment simulates a home and a shopping street in Japan (Figure\ref{fig:home_shopping_street}). 
Daily activities were performed from 6:30 AM to 10:30 PM with one hour equaling three minutes of real time. That is, a full day in virtual time was experienced in 48 minutes of real time. 
An analog clock was constantly displayed at the lower right of the participant's field of view, enabling participants to track the current time (Figure \ref{fig:VRT_clock_task} a)). 
The participants could execute a task while touching an object in the environment and correct the content on a choice menu (Figure \ref{fig:VRT_clock_task} b)).
The white box in Figure \ref{fig:VRT_clock_task} b) shows the English translation of the Japanese GUI.

Participants navigated between the home and shopping street, two typical settings in Japanese daily life. 
Referring to the clock, the participants were required to recall and execute given PM tasks at appropriate times. 
Executing a task involved moving through the VR environment and interacting with objects related to the task, triggering a dialog box with a button to confirm task completion. 
During the VRT phase, the participants completed both simple execution tasks and tasks requiring the selection of the correct action among multiple options.

Participants performed specified tasks within the accelerated time flow. The tasks were presented at the beginning of each VRT session.
% (Figure \ref{img:plan}). 
Regular tasks remained constant across all sessions, whereas irregular tasks varied from session to session. Each task was categorized as event-based (expected to be triggered by an event) or time-based (expected to be executed within a specified time window). 
Event-based tasks were judged solely on completion, whereas time-based tasks were evaluated on both completion and timing accuracy. The acceptable window was bounded by 15 virtual minutes before and 10 virtual minutes after the designated time.

Inspired by the Virtual Week task structure, the VRT phase included 10 PM tasks per day: five regular tasks (two time-based and three event-based) and up to five irregular tasks (three time-based and two event-based). 
This structure was designed to challenge participants with lower PM abilities (such as older adults) while being achievable by those with higher PM capabilities (such as university students). 
Considering the time required to navigate between the most distant points in the VR environment and perform the PM tasks, the task interval was set to one virtual hour (three minutes in real time).

The VR environment included 11 objects in the home and six objects in the shopping street for executing PM tasks. 
During the tutorial, participants watched a video and practiced regular tasks to familiarize themselves with the VR environment and the task-execution process.

The study also implemented two distractor tasks: a whack-a-mole game played in the home environment and a shooting gallery game played in the shopping street. 
Participants engaged in these background tasks until the optimal time of recall to execute their PM tasks. 
At the appropriate time, the participants moved to the designated location of the PM tasks within the VR environment.

\section{Experiment}

This experiment aimed to evaluate the effectiveness and usability of the VIT and VRT components of the VR--PMT system. The study ensured that all procedures adhered to the ethical guidelines of research involving human participants and was approved by the ethics committee of Kyoto University, Japan.
Participants were provided with a thorough explanation, both verbally and in writing, regarding the purpose of the study, methods, requirements, potential benefits and risks, and their rights to voluntary participation and withdrawal at any time.
Written informed consent was then obtained from all participants. Furthermore, the study ensured the protection of participants' privacy and personal information by anonymizing the data during its handling.

To assess the effectiveness of the system across different age groups, both university students and older adults were recruited for the study.
The participants were 10 healthy adults: five university students (mean age 22.4 $\pm$ 2.1 years) and five elderly participants (mean age 72.4 $\pm$ 8.0 years).
The experiment consisted of nine sessions conducted over nine weeks.
During the first session, the PM abilities of the participants were evaluated by a neuropsychology expert.
% Before the first session(Sessions 0), participants completed the Memory for Intentions Screening Test (MIST) to assess their baseline PM ability. 
The subsequent eight sessions involved training with the VR--PMT system as outlined in Table \ref{tab:schedule}.
Each training session was conducted once weekly.
Participants were allowed to take at least one break during each VRT session.
They could also request as many breaks as desired at any time and could terminate the session if necessary.

To determine whether our PM tasks truly reflect PM ability and to confirm the necessity of imagery, we computed the Pearson product-moment correlation coefficients.
Significant correlations between task performance and MIST scores would indicate that the tasks effectively measure the participants' PM abilities.
The correlation between task achievement and visual imagery checks whether our PM tasks are influenced by imagery abilities, emphasizing the need for imagery strategies.
Meanwhile, significant correlations between task performance and imagery task achievement would support the importance of incorporating imagery strategies into PM training.
To evaluate the design for difficulty progression, the task achievement rates of all participants in each session were determined. 
If the task completion rate does not decline in the latter half of the session, then it supports the appropriateness of the proposed design for difficulty progression.
Usability evaluation is a crucial components of system assessment.
Subjective fatigue was assessed on the Jikaku-sho Shirabe questionnaire, which focuses on work-related feelings of fatigue\cite{Jikaku_Sasaki_2008,Jikaku_Kubo_2011}.
The questionnaire comprises 25 subjective expressions in five categories: drowsiness, instability, uneasiness, dullness, and eyestrain.
The participants rated the intensities of their feelings on a five-point Likert scale, ranging from "totally disagree" to "strongly agree", for each item.
Scores of 1, 2, 3, 4, and 5 were assigned to intensity levels of 1, 2, 3, 4, and 5, respectively.
To assess the effect of age on fatigue during the VR training sessions, we compared the levels of reported fatigue between elderly and young participants across the five factors.

The task-related user experience was assessed on the User Experience Questionnaire (UEQ) \cite{UEQ_Laugwitz_2008, UEQ_Schrepp_2014, UEQ_Schrepp_2017}.
This study employed the short version of the UEQ (UEQ-S), which includes eight items related to efficiency, perspicuity, dependability, stimulation, and novelty\cite{UEQ_Laugwitz_2017}.
Participants rated their feelings for each item on a scale from -3 (most negative) to +3 (most positive).
To analyze the users' experience, the differences in ratings between elderly and young participants were compared to assess the perceived usability and enjoyment of the two groups.

% Our hypothesis are:
% \begin{enumerate}
% \item High correlation will be observed between MIST/imagery scores and task achievements of VRT.
% \item Event-based tasks has higher achievement than time-based one and regular tasks also has higher than irregular one.
% \end{enumerate}

\section{Result}

% The results indicate a significant correlation between both MIST scores and imagery abilities with task achievement across various task types, as shown in Table\ref{tab:correlation}. 
% For total task achievement, there is a very strong correlation with MIST scores ($r = 0.894, p < 0.0001$), indicating a highly significant relationship. 
% Regular tasks also show a strong correlation with MIST scores ($r = 0.787, p = 0.007$). 
% Irregular tasks display a very strong correlation with MIST scores ($r = 0.921, p = 0.0002$). 
% Event-based tasks have a strong correlation with MIST scores ($r = 0.729, p = 0.017$). 
% However, time-based tasks exhibit a very strong correlation with MIST scores ($r = 0.934, p = 0.0001$). 
% These findings suggest that the PM tasks implemented in the training system effectively reflect PM abilities, particularly for regular, irregular, and event-based tasks.

% Similarly, for total task achievement, there is a very strong correlation with imagery abilities ($r = 0.936, p < 0.0001$), indicating a significant relationship. 
% Regular tasks show a strong correlation with imagery abilities ($r = 0.907, p < 0.0001$). 
% Irregular tasks display a very strong correlation with imagery abilities ($r = 0.835, p = 0.003$). 
% Event-based tasks have a strong correlation with imagery abilities ($r = 0.787, p = 0.007$). 
% Time-based tasks also show a significant correlation with imagery abilities ($r = 0.964, p < 0.0001$). 

\begin{figure}[tbh!]
    \centering    \includegraphics[width=85mm]{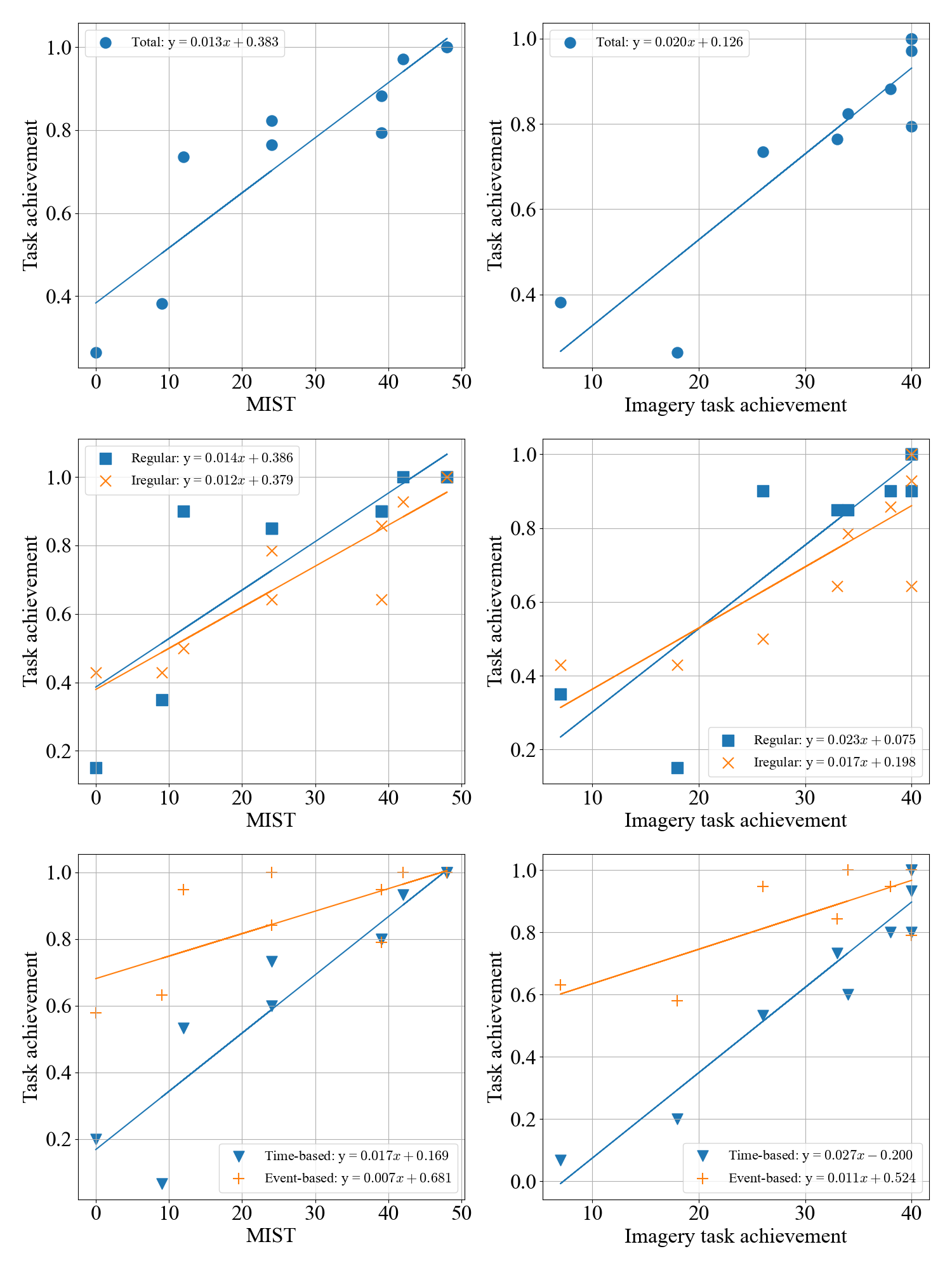}    \caption{Correlation analyses of total task achievement (top), achievements of regular and irregular tasks (center), and achievements of time-based and event-based tasks (bottom) vs. MIST and imagery abilities}
    \label{fig:correlation}\end{figure}

The scatter plots presented in Figure \ref{fig:correlation} show the correlations between task achievement as well as MIST scores and imagery task achievement across different task categories. 
The plots also include regression lines with corresponding equations to show the linear relationships in each case.
Additionally, the top-left plot demonstrates a positive correlation between overall task achievement and MIST scores, indicating an association between higher MIST scores and higher task achievement. 
Similarly, the top-right plot shows a strong correlation between overall task and imagery task achievements.
The middle row displays separate regression lines for regular and irregular tasks. 
Regular tasks exhibit a steeper slope than irregular tasks, indicating different trends in their relationships. 
The bottom row compares time- and event-based tasks, where the regression lines for time-based tasks show steeper slopes than those for event-based tasks.
% As shown in Table\ref{tab:correlation}, the MIST scores and imagery abilities were significantly correlated with task achievement across various task types. 
% Each row in this table presents the correlation coefficients ($r$), confidence intervals (95\% CI), and p-values ($p$), demonstrating the strength and significance of the ability-achievement relationships, respectively, in each task category.
% Both the MIST scores and imagery abilities were strongly or very strongly correlated with achievement in all task categories.
As shown in Table \ref{tab:correlation}, the MIST scores and imagery abilities were significantly correlated with task achievement across various task types. 
Each row presents the correlation coefficients ($r$), confidence intervals (95\% CI), and p-values ($p$), demonstrating the strength and significance of relationships between abilities and achievement in each task category. 
The MIST scores and imagery abilities exhibit strong correlations ($r > 0.6$) for event-based tasks and considerably strong correlations ($r > 0.8$) for total, regular/irregular, and time-based tasks. This highlights their influence on achievement across all categories.

% \begin{table}
% \centering
%     \caption{Operation on weighted feature detectors}
%     \label{tab:tabel_maxop}
%     \setlength{\tabcolsep}{3pt}
%     \begin{tabular}{p{100pt}p{25pt}p{25pt}}
%         \hline
%         \multicolumn{1}{c}{\multirow{2}{*}{Operation Mode}} & \multicolumn{2}{c}{mAP (\%)}\\
%         \cline{2-3}
%                             & Data1& Data2\\
%         Method1 (Original) & 32.5 & 16.9\\
%         Method1 (No1)       & 29.7 & 16.8\\
%         Method1 (No2)       & 32.1 & 17.0\\
%         Method1 (No3)       & 31.5 & 18.1\\
%         \hline
%     \end{tabular}
% \end{table}
% \begin{table}[tbh!]
%     \caption{Correlations of MIST and imagery scores with task achievement across different task types}
%     \centering   
%     \begin{tabular}{llllll}
%     \hline
%     Type     & \multicolumn{2}{c}{MIST} & & \multicolumn{2}{c}{Imagery} \\
%     \cline{2-3} \cline{5-6}
%              & $r$     & $p$                & & $r$     & $p$  \\ \hline
%     Total    & 0.894 & \verb|<| 0.0001      & & 0.936 & \verb|<| 0.0001   \\
%     Regular  & 0.787 & 0.007                & & 0.907 & \verb|<| 0.0001   \\
%     Irregular & 0.921 & \verb|<| 0.0001     & & 0.835 & 0.003  \\
%     Event    & 0.729 & 0.017                & & 0.787 & 0.007   \\
%     Time     & 0.934 & \verb|<| 0.0001      & & 0.964 & \verb|<| 0.0001  \\ \hline
%     \end{tabular}
%     \label{tab:correlation}
% \end{table}
\begin{table*}[tbh!]
    \caption{Correlations of MIST and imagery scores with task achievement across different task types}
    \centering   
    \begin{tabular}{llllllll}
    \hline
    Type     & \multicolumn{3}{c}{MIST} &                    & \multicolumn{3}{c}{Imagery} \\\cline{2-4} \cline{6-8}
             & $r$     & 95\% CI        & $p$                       & & $r$   & 95\% CI        & $p$  \\ \hline
    Total    & 0.910   & [0.656, 0.979] & \verb|<| 0.0001    & & 0.906 & [0.645, 0.978] & \verb|<| 0.0001   \\
    Regular  & 0.829   & [0.417, 0.958] & 0.003              & & 0.870 & [0.531, 0.969] & 0.001   \\
    Irregular & 0.921  & [0.693, 0.981] & \verb|<| 0.0001    & & 0.835 & [0.434, 0.960] & 0.003  \\
    Event    & 0.729   & [0.184, 0.931] & 0.017              & & 0.787 & [0.311, 0.947] & 0.007   \\
    Time     & 0.934   & [0.738, 0.985] & \verb|<| 0.0001    & & 0.964 & [0.849, 0.992] & \verb|<| 0.0001  \\ \hline
    \end{tabular}
    \label{tab:correlation}
\end{table*}

\begin{figure}[tbh!]    \centering
    \includegraphics[width=85mm]{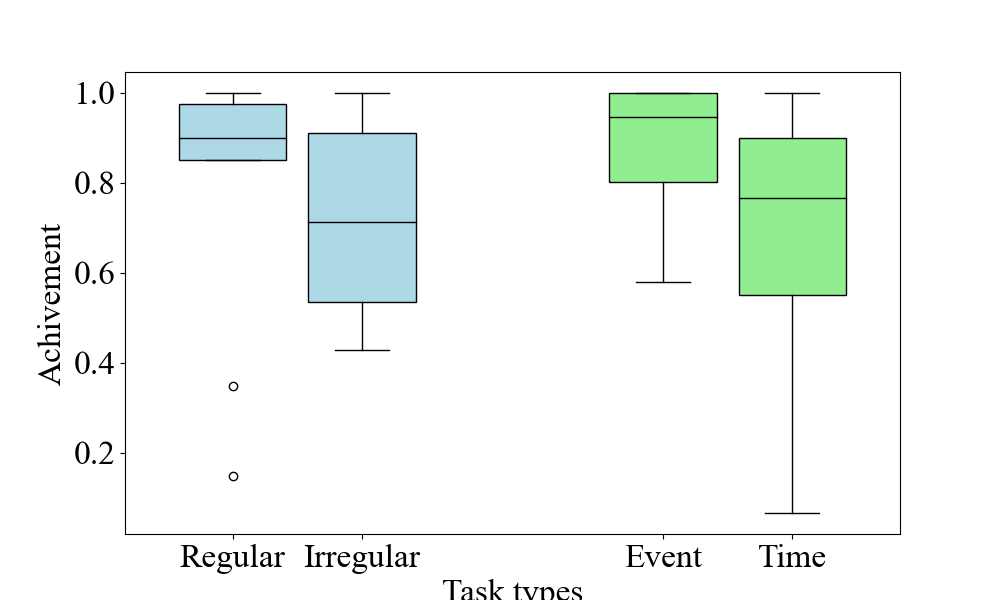}
    \caption{Comparison of task achievements across regular, irregular, event-based, and time-based tasks}
    \label{fig:Comparison_reg_event}
\end{figure}

Figure \ref{fig:Comparison_reg_event} visually compares the task achievements across the different task types. 
The box plots indicate that participants achieved higher scores in event-based tasks than in time-based tasks. They also obtained higher scores in regular tasks than in irregular tasks. 
This visual representation reinforces the statistical findings, highlighting the significant differences in task achievement based on task type and regularity.

Table \ref{tab:result_difficulty_progression} shows the task achievement rates for all participants across VRT sessions 5--8. 
The achievement rates for the elderly group (uppercase letters) were generally lower compared with those for the younger group (lowercase letters). 
A consistent decrease in the achievement rate was not observed across any participant groups. 

\begin{figure}[tbh!]
    \centering
    \includegraphics[width=80mm]{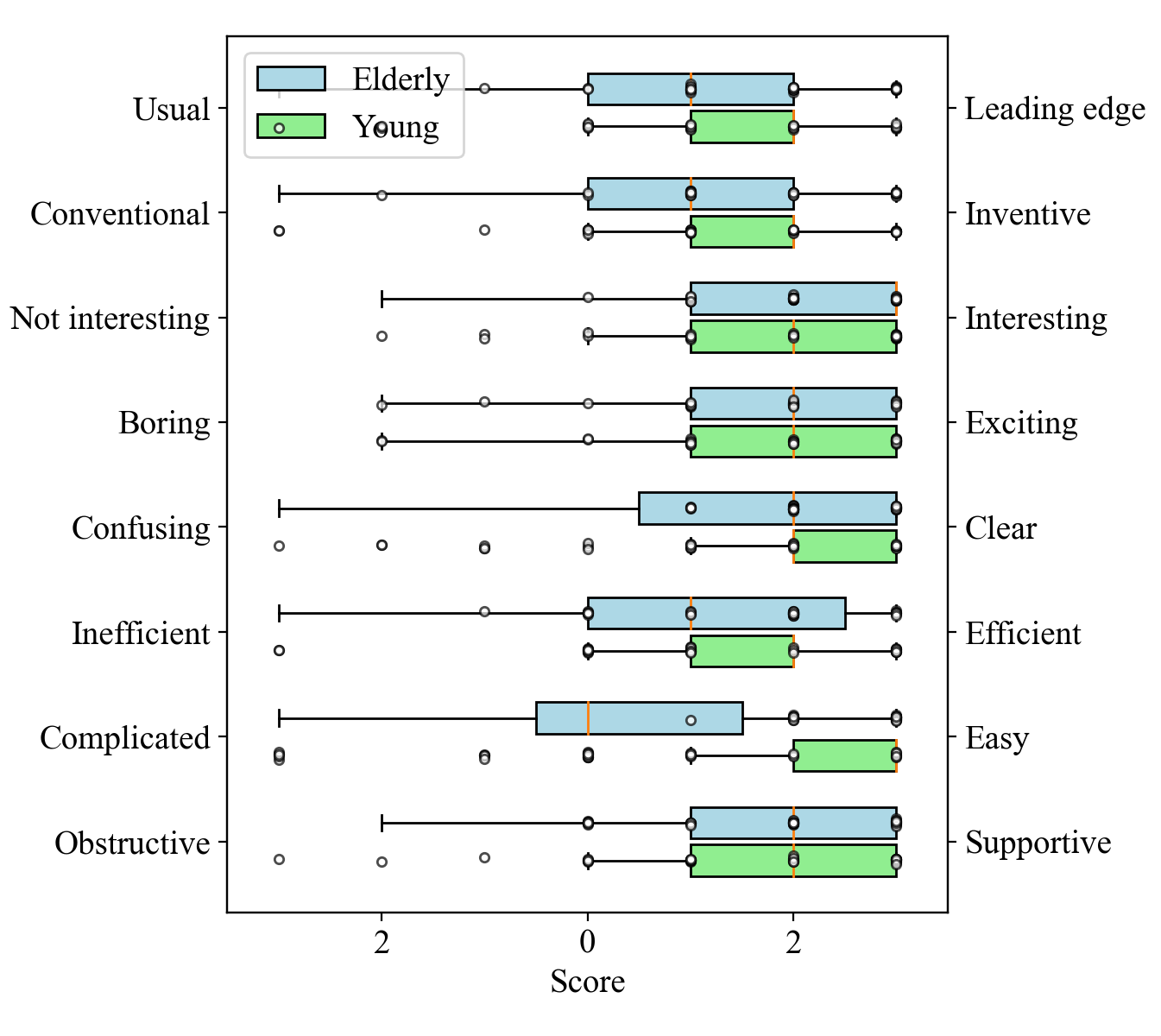}
    \caption{Comparison of user experience ratings (UEQ-S) between the elderly and young participants}
    \label{fig:UEQ}
\end{figure}

\begin{table}[]
    \caption{Task achievement rate in each session}
    \centering   
    \begin{tabular}{lllll}
    \hline
                & \multicolumn{4}{c}{Session} \\    \cline{2-5}
                & 5        & 6     & 7        & 8   \\
    Participant &          &       &          &     \\ \hline
    A           & 0.286 & 0.250 & 0.556 & 0.400 \\
    B           & 0.857 & 0.750 & 0.778 & 0.800 \\
    C           & 0.714 & 1.000 & 0.778 & 0.600 \\
    D           & 0.857 & 0.625 & 1.000 & 0.800 \\
    E           & 0.714 & 0.375 & 0.333 & 0.300 \\    
    A           & 1.000 & 1.000 & 1.000 & 1.000 \\
    b           & 1.000 & 1.000 & 0.889 & 1.000 \\
    c           & 1.000 & 1.000 & 1.000 & 1.000 \\
    d           & 1.000 & 0.750 & 1.000 & 0.500 \\
    e           & 0.857 & 0.875 & 1.000 & 0.800 \\ \hline
    \end{tabular}
    \label{tab:result_difficulty_progression}
\end{table}
% The UEQ results (Fig \ref{fig:UEQ}) demonstrate that both elderly and young participants rated the system positively across various aspects, indicating its overall effectiveness and appeal. 
% Elderly participants perceived the system as less "Leading edge," "Inventive," and "Interesting" compared to younger participants, but still rated it positively, suggesting that the system is considered innovative and engaging by all age groups.
% Young participants rated the system higher in terms of "Exciting" and "Clear," indicating that they found the system more stimulating and easier to understand compared to elderly participants. 
% Both groups gave positive ratings for "Efficient" and "Easy," with young participants providing higher scores. 
% Both elderly and young participants gave similar high ratings for "Supportive," suggesting that the system effectively aids users in completing their tasks. 
% The low scores for "Complicated," "Confusing," and "Obstructive" across both groups indicate that the system is not perceived as overly complex or difficult to use.
The UEQ-S results (Figure \ref{fig:UEQ}) demonstrate that both elderly and young participants rated the system positively across various satisfaction factors. 
However, the elderly participants perceived the system as less "Leading edge," "Inventive," and "Interesting" than younger participants. 
The system was rated as "Exciting" and "Clear" by the younger participants, "Efficient" and "Easy" by both groups (with young participants providing higher scores than their elderly counterparts), and "Supportive" by both groups. The low scores of "Complicated," "Confusing," and "Obstructive" across both groups indicate that the system is perceived as non-complex and reasonably easy to use.

\begin{figure}[tbh!]
    \centering
    \includegraphics[width=80mm]{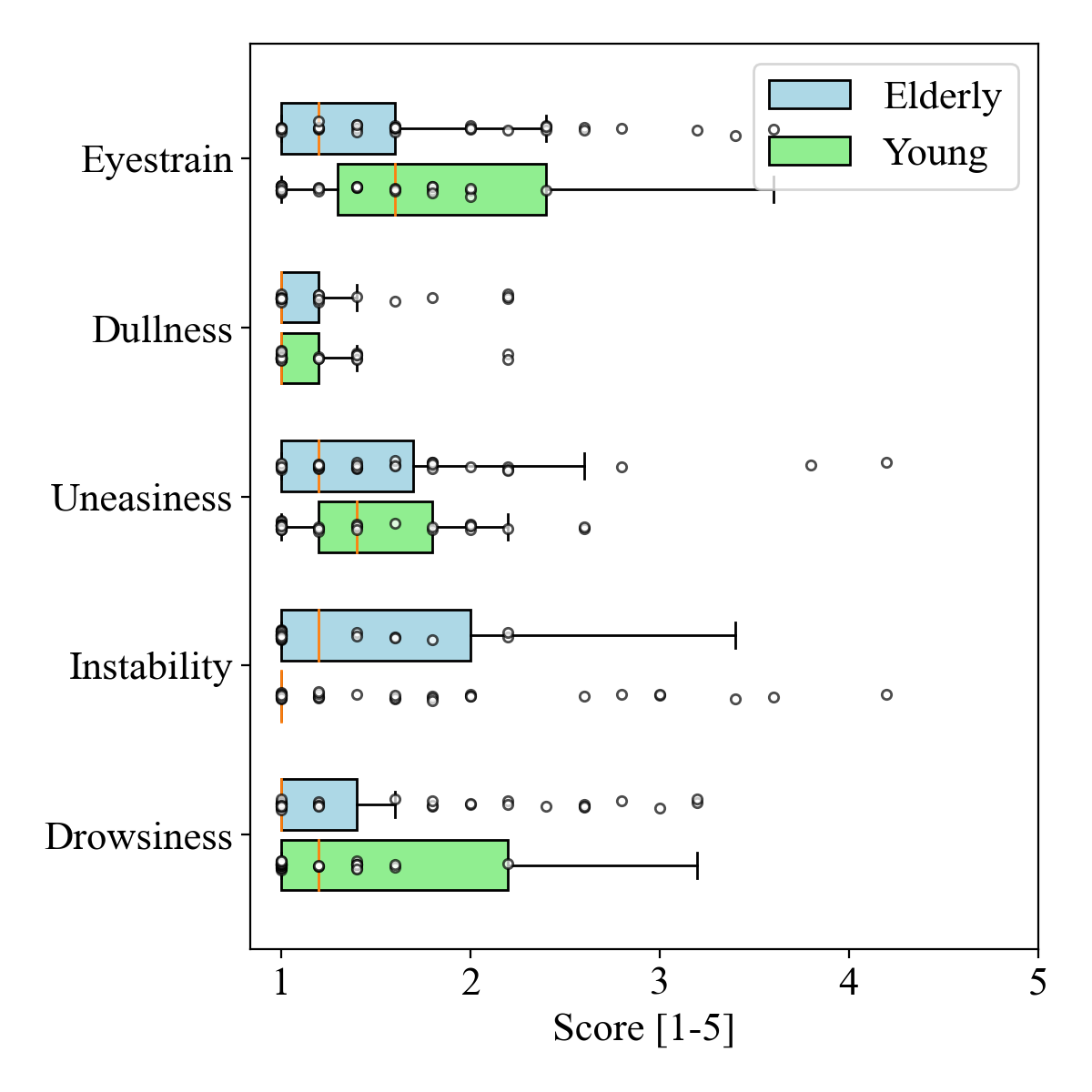}
    \caption{Comparison of subjective fatigue levels (Jikaku-sho Shirabe) between the elderly and young participants}
    \label{fig:Jikaku_sho_Shirabe}
\end{figure}

% The results(Fig\ref{fig:Jikaku_sho_Shirabe}) of the Jikaku-sho Shirabe questionnaire show different levels of fatigue between elderly and young participants.
% Specifically, elderly participants reported higher scores for eyestrain, dullness, uneasiness, instability, and drowsiness compared to younger participants.
% This suggests that elderly individuals tend to experience more fatigue during VR training.For eyestrain, elderly participants clearly showed higher scores than younger participants, indicating a greater visual burden for the elderly.
% Similarly, higher scores for dullness among the elderly suggest that overall fatigue is more pronounced in this group.
% In terms of uneasiness and instability, elderly participants again reported higher scores, reflecting a greater mental burden during VR training.
% Lastly, higher scores for drowsiness among the elderly indicate that longer training sessions may significantly increase fatigue for this age group.
The results (Figure \ref{fig:Jikaku_sho_Shirabe}) of the Jikaku-sho Shirabe questionnaire highlight varying levels of fatigue between the elderly and young participants. 
The elderly participants reported higher scores for eyestrain, dullness, uneasiness, instability, and drowsiness than younger participants, indicating
that elderly individuals were more fatigued by VR training then the younger participants.

\section{Post-hoc analysis: why is the correlation coefficient the lowest in event-based tasks?}

The timing of execution for event-based tasks is the primary focus of this study; event-based tasks exhibit distinct characteristics compared with other task types. 
This timing of execution is not linked to a specific time; thus, a participant can execute the task based on their preference. 
We hypothesize that this unique feature contributes to the lower correlation coefficient and wider confidence interval in event-based tasks compared with those in other task types.

Before conducting the post-hoc analysis to investigate the reasons for the lowest correlation coefficient in event-based tasks, all event-based tasks designed in the study were overviewed (Table \ref{tab:event-based_tasks}). 
Each task was identified by a unique ID for analysis, along with a description of the required action and the time-remembered task (TRT). 
The TRT indicates the duration at which the participants are expected to memorize the tasks. 
In the TRT column, "Before" indicates that the tasks were presented at the beginning of the VRT session, while specific times (e.g., 10:00 or 12:00) indicate that new tasks appeared at those designated times during the VRT session.
\begin{table}[tb]
    \caption{Detail of event-based tasks}
    \centering 
    \begin{tabular}{lp{4cm}ll}
    \hline
    ID  & Task                                                              & TRT      & Session\\    \hline
    ER1 & I will withdraw money from the ATM if there is a convenience store & Before   & 5\\
    ER2 & I will bring an umbrella when going out                           & Before   & 6\\
    ER3 & I will refill the shampoo when taking a bath                      & Before   & 7\\
    ER4 & I will repay the money I owe to Shimizu-san when I see him        & 10:00    & 7\\
    ER5 & I will pick up my suit from the dry cleaner when I go shopping    & Before   & 8\\
    ER6 & I will give the photos to Matsuda-san when I see him              & 12:00    & 8\\
    \hline
    \end{tabular}
    \label{tab:event-based_tasks}
\end{table}

Table \ref{tab:duration_event-based} shows the duration from when participants remembered a task to when they executed it for event-based tasks. It also shows the normalized values derived using the min--max method, enabling comparisons across different task types. The numeric values in each cell represent the time taken from remembering to executing a task, and the numbers within parentheses denote the corresponding normalized values in the one task. The table also distinguishes between elderly participants (denoted by uppercase letters) and younger participants (denoted by lowercase letters).
The normalized values (Table \ref{tab:duration_event-based}) show that participants B, c, and e frequently executed tasks much sooner than other participants; they completed four or more tasks in less than 20\% of the time taken by the slowest participant for the same task. 
Additionally, tasks ER4 and ER6 introduced in the middle of the VRT were shorter than other tasks.
\begin{table*}[tb]
    \caption{Duration from remembering the task to its execution and its normalized value obtained using the min--max method for each event-based task}
    \centering 
    \begin{tabular}{rrrrrrr}
        \hline
          & ER1           & ER2           & ER3           & ER4           & ER5           & ER6 \\
        Participant &     &               &               &               &  &  \\ \hline
        A & 08:05 (0.751) &               & 12:34 (1.000) & 04:47 (1.000) & 06:50 (0.429) & \\
        B & 02:40 (0.082) & 02:08 (0.003) & 02:19 (0.179) & & 07:32 (0.571) & 00:21 (0.022) \\
        C & 08:48 (0.840) & 03:54 (0.162) & 01:23 (0.104) & 02:50 (0.567) & 09:39 (1.000) & 00:44 (0.146) \\
        D & 04:29 (0.307) & 13:14 (1.000) & 12:21 (0.983) & 01:06 (0.181) & 05:31 (0.162) & 00:17 (0.000) \\
        E &               &               & 08:18 (0.658) & 02:59 (0.6)   & 04:43 (0.000) & \\
        a & 10:06 (1.000) & 08:20 (0.560) & 12:11 (0.969) & 02:58 (0.596) & 09:10 (0.902) & 03:22 (1.000) \\
        b & 08:15 (0.772) & 07:26 (0.479) & 12:10 (0.968) & 00:17 (0.000) & 08:45 (0.818) & 03:22 (1.000) \\
        c & 02:17 (0.035) & 02:42 (0.054) & 11:51 (0.943) & 00:27 (0.037) & 08:40 (0.801) & 01:45 (0.476) \\
        d & 08:07 (0.755) & 05:22 (0.293) & 09:40 (0.768) & 02:00 (0.381) & & 02:16 (0.643) \\
        e & 02:00 (0.000) & 02:06 (0.000) & 00:05 (0.000) & 00:36 (0.070) & 05:56 (0.247) & \\ \hline
        \end{tabular}
    \label{tab:duration_event-based}
\end{table*}

\section{Discussion}

As revealed in the scatter plots (Figure \ref{fig:correlation}), our VR training system accurately reflects the PM capabilities of the participants.
Specifically, individuals with higher MIST scores and better imagery task performance achieved higher task completion rates in the VRT tasks than those with lower scores.
Further analysis of the results shows that the VR training system effectively differentiates participants with varying PM abilities.
Moreover, the correlation between MIST scores and task achievement rates suggests that the system can reliably assess PM capabilities.
The implementation of visual imagery strategies within the VR environment appears to enhance task performance, as indicated by the positive relationship between imagery task achievement and overall task performance.
Besides providing a realistic training scenario, the integration of visual imagery strategies within the VR environment leverages the strengths of both VR and imagery-based approaches to enhance the training outcomes. These findings suggest that the proposed VR--PMT has the potential to assess PM capabilities.
The structure of the training program helps users to gradually build their confidence and skills, enabling adaptation to more challenging tasks over time. 
Figure \ref{fig:Comparison_reg_event} shows that the proposed task implementation in VRT was well designed, indicating  that time-based tasks are more difficult to undertake than event-based tasks. Moreover, irregular tasks were more difficult to undertake than regular ones.
% These findings are consistent with those reported by Kant\cite{achievement_time<event}, which revealed lower achievement rates for time-based task than event-based tasks.
The findings are also consistent with those reported by Rendell\cite{achievement_irreg<reg}, which revealed that the user performance on irregular tasks was lower than on regular tasks.
This confirmed the consistency of our design intention with task difficulty.
The difficulty structure in the progression of training program began with simple tasks (regular and event-based) and progresses to more complex tasks (irregular and time-based) as the sessions proceed. Table \ref{tab:result_difficulty_progression} indicates that there were no significant decrements in task achievement rates across sessions and that the difficulty progression in the training design was generally appropriate.
The rate of task achievements in elderly participants was more variable than young participants. However, for most participants, including younger and older ones, the task achievement rates remained relatively stable, even when the difficulty level increased. 
This stability indicates that the gradual increase in complexity did not destabilize the user performance and allowed them to maintain consistent performance throughout. 
The absence of significant decrements in task completion rates, particularly in the latter sessions, supports the effectiveness of the training design, which aimed to progressively increase the task difficulty.

% The reason for different achievement rates of event-based task and other tasks must be elucidated to improve the performance of the proposed framework.
% As shown in Table \ref{tab:duration_event-based}, event-based tasks have the lowest correlation coefficients, the widest confidence intervals, and the flattest regression slopes compared with other task types.

The proposed framework effectively simulated realistic situations, allowing participants to execute some tasks according to their own strategies (Table \ref{tab:duration_event-based}). 
Table \ref{tab:duration_event-based} shows that some participants (B, c, and e) tended to complete their tasks as quickly as possible because they executed the tasks immediately after remembering them. This tendency likely contributed to higher PM achievement rates.
Thus, whether the event-based tasks in this framework contribute to the improvement of prospective memory abilities requires further analysis. However, the framework effectively learns how to execute planned actions. 
If the planned actions are independent of other tasks or specific times, then completing them as early as possible is a reasonable behavior in everyday life for training aimed at achieving daily goals. 
Therefore, the framework enables participants to recognize their own capabilities and learn practical strategies for early execution of tasks.
Acquiring such strategies is useful during the course of training, as it allows participants to develop strategies suited to their abilities.

The unique timing characteristics of event-based tasks were studied to determine the difference in correlation patterns compared with those of other task types. 
Unlike that for time-based tasks, the execution timing for event-based tasks is flexible. This may account for the observed lower correlation coefficients, wider confidence intervals, and flatter regression slopes in Figure \ref{fig:correlation}.
The flexibility in timing allows participants to employ various strategies for executing tasks. 
For instance, the normalized values in Table \ref{tab:duration_event-based} show that participants B, c, and e often executed tasks shortly after remembering them. 
This immediate execution suggests a strategy for minimizing the load toward PM by quickly completing tasks for avoiding the possibility of forgetting them. 
Such an approach can reduce the participants’ dependency on PM abilities because they may not need to retain task information for longer durations and instead reinforce task recall through rapid memory rehearsal.
This strategy explains the weaker relationship between PM scores and task achievement for event-based tasks. 
By executing tasks immediately, participants effectively reduce the influence of their underlying PM capabilities on task performance. Consequently, the variability in PM abilities among participants does not manifest as distinctly in their task achievement, which is clearly reflected in the flatter regression slopes and weaker correlations.

Whether event-based tasks in the proposed framework improve the PM abilities requires further investigation. 
For effective training, the design for event-based task should be improved to enhance its influence on PM ability and better simulate real-world complexities. This will prevent participants from executing tasks immediately after remembering them.
One potential improvement is to present an event-based task to participants shortly before they are scheduled to perform a time-based task. Since the time-based task has a fixed execution time, participants would need to prioritize completing it first, delaying the execution of the event-based task.  However, the framework is effective in terms of learning how to execute planned actions. 
For training aimed at achieving daily goals, if the planned actions are independent of other tasks or specific times, then completing them as early as possible is a reasonable behavior in everyday life. 
Therefore, the framework enables participants to recognize their own capabilities and learn practical strategies for executing tasks earlier.
Acquiring such strategies is useful during the course of training as it allows participants to develop strategies suited to their individual abilities.% ----------------------------------------------------------------
% UEQ-S 
% ----------------------------------------------------------------
The VR-PMT platform was considered as innovative and engaging by both age groups (Figure \ref{fig:UEQ}). 
The higher ratings of "Exciting" and "Clear" by younger participants indicate that these participants found the system more stimulating and understandable than the elderly participants. 
The similarly high ratings of "Supportive" by both age groups suggest that the system effectively aids users in completing their tasks. 
The low scores of "Complicated," "Confusing," and "Obstructive" indicate the user-friendliness and relative simplicity of the system.
% ----------------------------------------------------------------

% ----------------------------------------------------------------
% Jikaku-sho Shirabe 
% ----------------------------------------------------------------
As indicated in Figure \ref{fig:Jikaku_sho_Shirabe}, the VR-PMT platform can potentially provide a stimulating and engaging training environment while pinpointing areas that should be improved to enhance user comfort and reduce fatigue.
The elevated eyestrain scores suggest that VR imposes a high visual demand, potentially enhancing the sense of immersion and focus of elderly users. 
The higher dullness and uneasiness scores might reflect the intense cognitive engagement required by VR training, indicating that the system effectively stimulates mental activity. 
The increased instability and drowsiness scores of the elderly participants than of the younger participants highlight the need for tailored interventions, such as regular breaks and ergonomic adjustments, to optimize the training experience of older users. % ----------------------------------------------------------------

Despite being a preliminary study with a limited sample size, this study can contribute to VR-based PM training. 
Moreover, this study primarily introduced a more detailed framework for task design and implementation compared with those reported in previous studies, providing in-depth guidance for developing a VR-based PM training system. 
The task structures were outlined, and the framework design was validated. The task achievement rates reflected the PM abilities of participants that were consistent with the findings from existing literature.
Although the confidence intervals for correlation coefficients were relatively narrow, which supported the reliability of the observed relationships across most task types, event-based tasks demonstrated a different pattern. 
This distinct trend was thoroughly examined by analyzing the task execution timing in detail. Results provided insights into the participants' strategies that had not been previously reported in studies related to VR-based PM training. 
By addressing this variability, we provided a novel understanding of the influence of execution timing on the PM task performance and laid the groundwork for future improvements in training design.
\EOD

\end{document}